\documentclass[aps,twocolumn, showpacs,preprintnumbers,amsmath,amssymb,prl,superscriptaddress]{revtex4-2}

\usepackage{xr}

\usepackage{graphicx}
\usepackage{dcolumn}
\usepackage{bm}
\usepackage{hyperref}
\usepackage{CJK}
\usepackage{amsmath,amssymb,bbm}
\graphicspath{{./figures}}

\usepackage{tikz,xcolor}
\usepackage{pgffor}  
\usepackage{avant}   

\definecolor{lime}{HTML}{A6CE39}
\DeclareRobustCommand{\orcidicon}{
	\begin{tikzpicture}
	\draw[lime, fill=lime] (0,0) 
	circle [radius=0.16] 
	node[white] {{\fontfamily{qag}\selectfont \tiny ID}};
	\draw[white, fill=white] (-0.0625,0.095) 
	circle [radius=0.007];
	\end{tikzpicture}
	\hspace{-2mm}
}

\foreach \x in {A, ..., Z}{%
	\expandafter\xdef\csname orcid\x\endcsname{\noexpand\href{https://orcid.org/\csname orcidauthor\x\endcsname}{\noexpand\orcidicon}}
}

\usepackage[normalem]{ulem}
\newcommand{\rsout}{\bgroup \color{red} \ULdepth=-.5ex \ULset}

\newcommand{\vct}[1]{{\boldsymbol{#1}}}

\begin{document}
\begin{CJK*} {UTF8} {gbsn}




\title{Resonance Femtoscopy Beyond the On-Shell Approximation}


\author{Liang Zhang(张良)\orcidA{}}
\affiliation{Key Laboratory of Nuclear Physics and Ion-beam Application (MOE), Institute of Modern Physics, Fudan University, Shanghai 200433, China}
\author{Tianhao Shao(邵天浩)\orcidB{}}
\affiliation{Key Laboratory of Nuclear Physics and Ion-beam Application (MOE), Institute of Modern Physics, Fudan University, Shanghai 200433, China}
\author{Song Zhang(张松)\orcidC{}}
\email{song\_zhang@fudan.edu.cn}
\affiliation{Key Laboratory of Nuclear Physics and Ion-beam Application (MOE), Institute of Modern Physics, Fudan University, Shanghai 200433, China}
\affiliation{Shanghai Research Center for Theoretical Nuclear Physics, NSFC and Fudan University, Shanghai 200438, China}
\author{Kai-Jia Sun(孙开佳)\orcidD{}}
\email{kjsun@fudan.edu.cn}
\affiliation{Key Laboratory of Nuclear Physics and Ion-beam Application (MOE), Institute of Modern Physics, Fudan University, Shanghai 200433, China}
\affiliation{Shanghai Research Center for Theoretical Nuclear Physics, NSFC and Fudan University, Shanghai 200438, China}
\author{Yu-Gang Ma(马余刚)\orcidE{}}
\email{mayugang@fudan.edu.cn}
\affiliation{Key Laboratory of Nuclear Physics and Ion-beam Application (MOE), Institute of Modern Physics, Fudan University, Shanghai 200433, China}
\affiliation{Shanghai Research Center for Theoretical Nuclear Physics, NSFC and Fudan University, Shanghai 200438, China}
\affiliation{School of Physics, East China Normal University, Shanghai 200241, China}


\date{\today}

\begin{abstract}
The observed shift of the $\Delta(1232)$ resonance peak in $\pi$-$p$ femtoscopic correlations challenges the conventional Breit-Wigner description of resonances in femtoscopy. 
We revisit the Koonin-Pratt framework by formulating femtoscopy in the momentum-space representation. By employing the T-matrix approach to disentangle on-shell and off-shell contributions, we show that the finite spatial extent of the emission source naturally induces sensitivity to off-shell scattering dynamics.
Using a Friedrichs-Lee model constrained by low-energy $\pi$-$p$ scattering data, we numerically demonstrate that this off-shell sensitivity leads to a peak shift accompanied by a dip on the high-momentum side of the peak. 
The predicted high-momentum side dip is absent in the data, pointing to source properties beyond the simple Gaussian approximation.
\end{abstract}


\maketitle

\emph{Introduction.}{\bf ---}
Femtoscopic correlations~\cite{Lisa:2005dd,Wiedemann:1999qn,Fabbietti:2020bfg} in high-energy nucleus-nucleus collisions are governed by both the properties of the particle emission source~\cite{Kopylov:1974th,Koonin:1977fh,Zajc:1984vb,Pratt:1986cc,Chapman:1995nz,Wiedemann:1999qn,ALICE:2023sjd} and the quantum evolution of the emitted pair~\cite{HanburyBrown:1956bqd,Goldhaber:1960sf,Pratt:1986ev,lednicky_influence_1996,lednicky_femtoscopy_2001,Lednicky:2002fq,MaYG:PRC06,STAR:2015kha,FangDQ:PRC16,Kamiya:2019uiw,ALICE:2019gcn,ALICE:2020mfd,ALICE:2021cpv,WangTT:PRC23,WangTT:PRC24,DingMQ:NST,XiBS:NST,STAR:dL}. 
In this sense, femtoscopy resembles a ``binary equation'' 
: a single measured correlation function must constrain two intrinsically entangled unknowns, the source properties and the pair evolution.
 

Recently, the ALICE Collaboration measured the $\pi$-$p$~\cite{ALICE:2025aur} and $\pi$-$d$~\cite{ALICE:2025byl} correlation functions in $\sqrt{s}=13~\rm{TeV}$ $pp$ collisions, revealing that the extracted $\Delta$-resonance peak position deviates significantly from the value reported by the Particle Data Group (PDG)~\cite{ParticleDataGroup:2024cfk,Hunt:2018wqz}.
This observation is of particular importance, as it challenges the conventional assumption that resonance properties entering femtoscopic analyses can be directly taken from vacuum scattering data.
The shift has been interpreted as a very low spectrum temperature~\cite{ALICE:2025byl,ALICE:2025aur}. It may also be caused by in-medium modification of the $\Delta$ mass at high temperature~\cite{Zhang:2025tfd}, due to the temperature dependence of decuplet baryons~\cite{Xu:2015jxa,vanHees:2004vt}.
A similar deviation was also reported by the HADES Collaboration in $\sqrt{s_{NN}}=2.42~\rm{GeV}$ Au+Au collisions~\cite{Adamczewski-Musch:2020edy}, where it was attributed to $\pi$-$p$ re-scattering effects in a thermal medium prior to kinetic freeze-out~\cite{Reichert:2019lny}.

Despite their different physical pictures, these interpretations share a common feature: they attribute the observed shift entirely to modifications of the emission source, i.e., the space-time structure and late-stage hadronic interactions encoded in the source function.
This raises a fundamental challenge for femtoscopy, namely whether the observed correlation signal can be unambiguously factorized into a ``source'' component and a ``final-state interaction'' kernel when resonant channels are present.
In particular, the $\Delta$ resonance, due to its short lifetime and strong coupling to the $\pi$-$p$ channel, not only modifies the emission history but also governs the quantum evolution of the two-particle wave function itself.
Neglecting this dual role may lead to an incomplete—or even biased—interpretation of the correlation signal.
A consistent description therefore requires going beyond source-level modeling and incorporating the $\Delta$ dynamics explicitly within a scattering-theory framework, where resonance formation, propagation, and decay are treated on equal footing with the femtoscopic correlation.
This implies that the observed peak shift is not merely a probe of the emission source, but a direct manifestation of the breakdown of the conventional on-shell source-kernel factorization in the presence of strong resonant interactions.

In scattering theory, the evolution of the emitted pair is encoded in the full scattering T-matrix. For the $\pi$-$p$ system in the energy range of interest, the interaction is dominated by a single $\Delta$ resonance pole in the $P_{33}$ channel, making it an ideal benchmark for an analysis based on T-matrix.
The T-matrix can be tightly constrained by low-energy $\pi$-$p$ scattering data, providing a data-driven baseline for the pair-evolution sector of the femtoscopic ``binary equation'' and leaving the source properties as the primary unknown to be extracted.

While off-shell contributions to femtoscopic correlations have been recognized in recent studies~\cite{Epelbaum:2025aan,Xie:2026hpp}, this motivates a treatment of resonance femtoscopy based on the full T-matrix, which naturally encodes the off-shell dynamics.
To this end, we incorporate a general T-matrix formulation into the Koonin-Pratt framework~\cite{Koonin:1977fh,Pratt:1986ev} to establish a scattering-based baseline for interpreting resonance structures in femtoscopic observables.
In this approach, the on-shell and off-shell contributions can be cleanly separated through a spectral decomposition of the correlation function, allowing us to trace the origin of the resonance peak in the established baseline.

The primary goal of the present work is to establish a scattering-constrained baseline for resonance femtoscopy and to analyze the respective roles of on-shell and off-shell contributions to the correlation function.
To implement this framework quantitatively, we employ a Friedrichs-Lee model~\cite{friedrichs_perturbation_1948,Lee:1954iq} constrained by low-energy $\pi$-$p$ scattering data~\cite{Carter:1971tj} to construct a quantitative reference framework for femtoscopic correlations in high-energy collisions. 
Before invoking complex in-medium modifications, this framework, combined with a simple Gaussian source, naturally reproduces the characteristic width and peak position of the observed correlation structure in experiment. 
The remaining discrepancy encodes genuine information about source properties beyond the simple Gaussian approximation.

\emph{Femtoscopic correlations in momentum space}.{\bf ---}\label{sec:Femto}
The Koonin-Pratt formula~\cite{Koonin:1977fh,Pratt:1986ev} can be expressed in Dirac notation,

\begin{equation}
\label{eq:KP}
\begin{aligned}
     C(\vct{p})=&\int d^3\vct{k}d^3\vct{k'}\langle \psi^{(+)}_\vct{p}|\vct{k}\rangle\langle\vct{k}|\hat{S}|\vct{k}'\rangle\langle\vct{k}'|\psi^{(+)}_\vct{p}\rangle
\end{aligned}
\end{equation}
where $\hat{S}$ represents the normalized emission source, 
and $|\psi^{(+)}_\vct{p}\rangle$ is the scattering state satisfying the outgoing boundary condition. 
Although the emission source for target pair $\langle \vct{r}|\hat{S}|\vct{r}'\rangle$ is conventionally approximated as diagonal in coordinate space, a more fundamental observation is that any source of finite spatial extent necessarily acquires a non-diagonal structure in the momentum representation under Fourier transform.
This non-diagonal structure reflects the ability of the source to couple different momentum states. 
When inserted into the Koonin-Pratt formula [Eq.~\eqref{eq:KP}] together with the scattering wave function, this non-diagonality probes momentum configurations with $k \neq p$, thereby granting access to the off-shell sector of the scattering dynamics.
As will be shown below, such off-shell contributions play an essential role in understanding the experimentally observed peak displacement.

The scattering wave function $\langle\vct{k}|\psi^{(+)}_\vct{p}\rangle$ is constructed from the T-matrix via the Lippmann-Schwinger equation,
$\langle \vct{k}|\psi^{(+)}_\vct{p}\rangle = 
\delta^{(3)}(\vct{k}-\vct{p}) + G^{(+)}(\vct{k};E)T(\vct{k},\vct{p};E)$,
where $G^{(+)}(\vct{k};E) = 2\mu/(p^2-k^2+i\epsilon)$ is the continuum propagator, with $E$ and $p=|\vct{p}|$ being the eigenenergy and momentum of $|\psi_{\vct{p}}^{(+)}\rangle$.
In the energy range considered here, only a single resonance contributes, and the T-matrix can be generally expressed as $T(\vct{k},\vct{p};E)=g(\vct{k})g^*(\vct{p})/(E-E_{0}-\Sigma_0(E)+i\epsilon)$~\cite{Yamaguchi:1954mp,Yamaguchi:1954zz}, where the spin projector $\mathcal{P}_{J_R}$ is kept implicit in the 
vertex function $g(\vct{k})$.
The pole in the complex-energy plane and the resonance peak position on the real axis are determined by the real bare mass $E_{0}$ and the complex self-energy $\Sigma_0(E)$. Unitarity requires that the self-energy satisfy $\operatorname{Im}\Sigma_0(E)=-\pi\rho(E)|g(\vct{p})|^2$, where $\rho(E)$ is the density of states at energy $E$ and $\vct{p}$ is the on-shell momentum.

Substituting the scattering wave function into the Koonin-Pratt formula [Eq.~\eqref{eq:KP}], the correlation function naturally decomposes into three components,
\begin{equation}
\label{eq:LS+KP}
\begin{aligned}
    C(\vct{p})=&1+2\operatorname{Re}\langle \vct{p}|\hat{S} G^{(+)}T|\vct{p}\rangle+
    \langle \vct{p}|T^\dagger {G^{(+)}}^\dagger\hat{S} G^{(+)}T|\vct{p}\rangle\\
    =&1+C_{\text{int}}(p)+C_{\text{scat}}(p),
\end{aligned}
\end{equation}
in which the $C_{\text{int}}$ term represents the interference between the incident wave and the scattered wave, and $C_{\text{scat}}$ accounts for the pure scattering contribution to the femtoscopic correlation. 
Both terms hinge on the continuum propagator $G^{(+)}(\vct{k};E_p)$, 
whose decomposition is the central step of the analysis.
By applying the Sokhotski-Plemelj theorem, we perform a ``spectral surgery'' on the propagator into a principal value part and a pole contribution,
\begin{equation}
\label{eq:Sokhotski-Plemelj}
    \frac{2\mu}{ p^2-k^2\pm i\epsilon}=\text{P.V.}\frac{2\mu}{ p^2-k^2}\mp i2\mu\pi\delta(p^2-k^2).
\end{equation}
After isolating the on-shell contribution associated with the $\delta$ function in the propagator, the remaining principal-value integral probes momentum configurations away from the pole at $k=p$. In this sense, it encodes the off-shell dynamics weighted by the source function.
This decomposition serves as a diagnostic tool to disentangle different contributions to the correlation function.

Carrying out the momentum integrals with the $\delta$-function part gives the on-shell contributions, 
\begin{equation}
\begin{aligned}
\label{eq:Cint-on-shell}
C^{\text{(on-shell)}}_{\text{int}}&(\vct{p})\\
=-&\frac{\Gamma(E)^2/2}
{[E-M_R(E)]^2+\Gamma(E)^2/4}(S\mathcal{P}_{J_R})(\vct{p},\vct{p}),    
\end{aligned}
\end{equation}
\begin{equation}
\begin{aligned}
\label{eq:Cscat-on-shell}
C^{\text{(on-shell)}}_{\text{scat}}&(\vct{p})\\
=&\frac{\Gamma(E)^2/4}
{[E-M_R(E)]^2+\Gamma(E)^2/4}(\mathcal{P}^\dagger_{J_R}S\mathcal{P}_{J_R})(\vct{p},\vct{p}).    
\end{aligned}
\end{equation}
Here the resonance mass and width are defined as 
$M_R(E)=E_0+\mathrm{Re}[\Sigma_0(E)]$ and
$\Gamma(E)=-2\,\mathrm{Im}[\Sigma_0(E)]$, respectively.
With $|g|^2$ traded for $\Gamma$ via the unitarity relation $\operatorname{Im}\Sigma_0=-\pi\rho|g|^2$, the projector $\mathcal{P}_{J_R}$ emerges as the remaining spin structure from the vertex product $g(\vct{k})g^*(\vct{p})$, which projects onto the total spin-$J_R$ resonance state.
The on-shell parts of $C_{\text{int}}(p)$ and $C_{\text{scat}}(p)$ both exhibit Breit-Wigner profiles~\cite{Breit:1936zzb,Kamiya:2019uiw}. Crucially, the two contributions appear with opposite signs: $C_{\text{int}}$ is negative and twice as large in magnitude as the positive $C_{\text{scat}}$. 
Therefore, in the present framework, the on-shell sum leads to a net dip, and the resulting peak in the correlation function emerges predominantly from off-shell dynamics encoded in the principal-value integral of the propagator.

\begin{figure}
    \centering
    \includegraphics[width=0.95\linewidth]{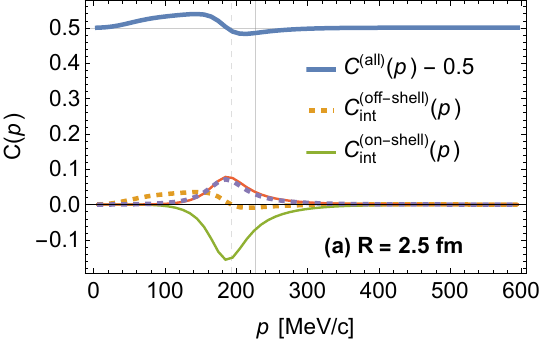}
    \\[2ex]
    \includegraphics[width=0.95\linewidth]{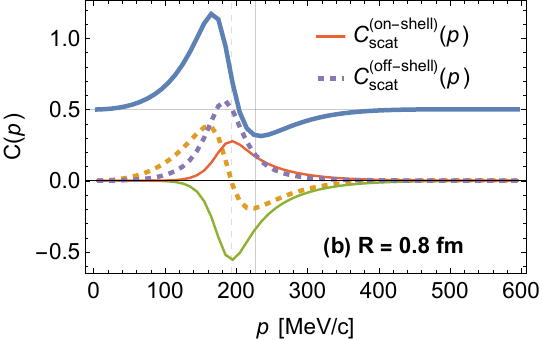}
    \caption{Resonance contributions to the $\pi$-$N$ correlation function (without Coulomb contribution), showing that the peak is dominated by off-shell dynamics. Results are shown for spherically symmetric Gaussian sources with radii of 2.5 fm (a) and 0.8 fm (b). The T matrix is obtained from the Friedrichs-Lee model, as described in the resonance model section. 
    The gray vertical solid line indicates the $\Delta$ Breit-Wigner mass from the PDG~\cite{ParticleDataGroup:2024cfk,Hunt:2018wqz} using the relativistic dispersion relation, while the dashed line indicates the non-relativistic dispersion relation.}
    \label{fig:Corr-Apart}
\end{figure}

The off-shell contribution of $C_{\text{int}}$, from the principal-value integration, introduces a dispersive structure,
\begin{equation}
\begin{aligned}
\label{eq:Cint-offshell}
    C^{\text{(off-shell)}}_{\text{int}}&(\vct{p})\\
    =&\frac{M_R(E)-E}{[E-M_R(E)]^2+\Gamma(E)^2/4}\times\mathcal{F}_{\text{int}}(E, S)
\end{aligned}
\end{equation}
where the source response factor $\mathcal{F}_{\text{int}}(E, S)$ is a smooth, positive function that encodes the interplay between the source geometry and the T-matrix vertices. The sign of $C^{\text{(off-shell)}}_{\text{int}}$ is governed entirely by the prefactor $M_R(E)-E$, which is positive below the resonance and negative above it, producing the characteristic peak-dip asymmetry (Fig.~\ref{fig:Corr-Apart}).

The off-shell parts of $C_{\text{scat}}$ also provide a contribution from a skewed Breit-Wigner peak which is strongly influenced by a second positive response factor $\mathcal{F}_{\text{scat}}(E, S)$,
\begin{equation}
\begin{aligned}
\label{eq:Cscat-offshell}
    C^{\text{(off-shell)}}_{\text{scat}}&(\vct{p})\\
=&\frac{\Gamma(E)/[2\pi \rho(E)]}{[E-M_R(E)]^2+\Gamma(E)^2/4}\times \mathcal{F}_{\text{scat}}(E, S).
\end{aligned}
\end{equation}
Due to the cancellation between the on-shell part of $C_{\text{scat}}(p)$ and $C_{\text{int}}(p)$, the off-shell contribution of $C_{\text{scat}}(p)$ becomes dominant (Fig.~\ref{fig:Corr-Apart}). This results in the characteristic peak shift toward lower relative momentum, which is a reflection of off-shell dynamics. The explicit expressions for $\mathcal{F}_{\text{int}}(E, S)$ and $\mathcal{F}_{\text{scat}}(E, S)$ are given in Eqs.~(S21)-(S22) of the Supplemental Material.

\emph{The resonance model}.{\bf ---}\label{sec:models}
We model the $\pi$-$N$ interaction using the Friedrichs-Lee framework~\cite{friedrichs_perturbation_1948,Lee:1954iq,Xiao:2023lpv}, in which a bare $|\Delta\rangle$ state couples to the $\pi$-$N$ scattering continuum via a $P$-wave vertex. The $\Delta$ resonance then emerges as a dressed pole in the $P_{33}$ T-matrix, rather than as an elementary particle. Although relativistic corrections become noticeable for $p\gtrsim 100$~MeV/$c$, we adopt a non-relativistic approximation, which does not affect the qualitative shape of the correlation function.

\begin{figure}
    \centering
    \includegraphics[width=0.95\linewidth]{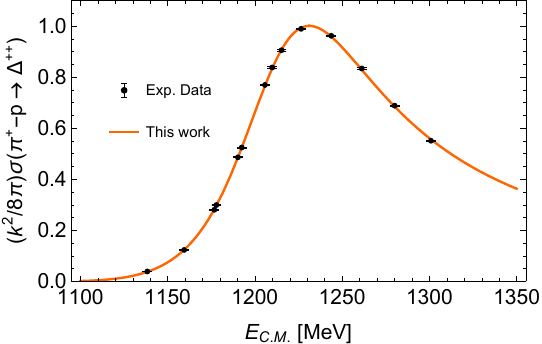}
    \caption{Resonance cross section of $\pi^+-p\rightarrow\Delta^{++}$ after background subtraction~\cite{Carter:1971tj,Haskins:1985xg,Giacosa:2021mbz}, fitted using the Friedrichs-Lee model. 
    }
    \label{fig:FitCrosssection}
\end{figure}

The $P_{33}$ interaction is parameterized by a $P$-wave vertex $g(k) \propto k f(k)$ with a dipole form factor $f(k) = (1+k^2/\Lambda^2)^{-2}$~\cite{Dominguez:1980jv,Gridnev:2018rbw} that regulates the ultraviolet behavior. The resulting T-matrix takes the separable form
\begin{equation}
\label{eq:T32matrix}
    T_{3/2}(k',k;E)=\frac{\left(\frac{g_0}{m_\pi}\right)^2 k'k f(k)f(k')} {E-E_{\Delta0}-\Sigma_0(E)+i\epsilon},
\end{equation}
where the self-energy $\Sigma_0(E)$ encodes the dressing of the bare state by the continuum. The two free parameters $(g_0,\Lambda)$, together with the bare mass $E_{\Delta0}$, are fixed by fitting the background-subtracted $\pi^+p\to\Delta^{++}$ cross section extracted from Ref.~\cite{Carter:1971tj,Haskins:1985xg,Giacosa:2021mbz} (Fig.~\ref{fig:FitCrosssection}). The fitted T-matrix reproduces the vacuum $\Delta(1232)$ mass and width~\cite{ParticleDataGroup:2024cfk,Hoferichter:2023mgya,Hunt:2018wqz}, ensuring that our subsequent femtoscopic analysis is anchored to experimentally constrained scattering dynamics. The detailed Hamiltonian, fitting procedure, and parameter values are provided in the Sec.~S3 of Supplemental Material.

\begin{figure*}
    \centering
    \includegraphics[width=1\linewidth]{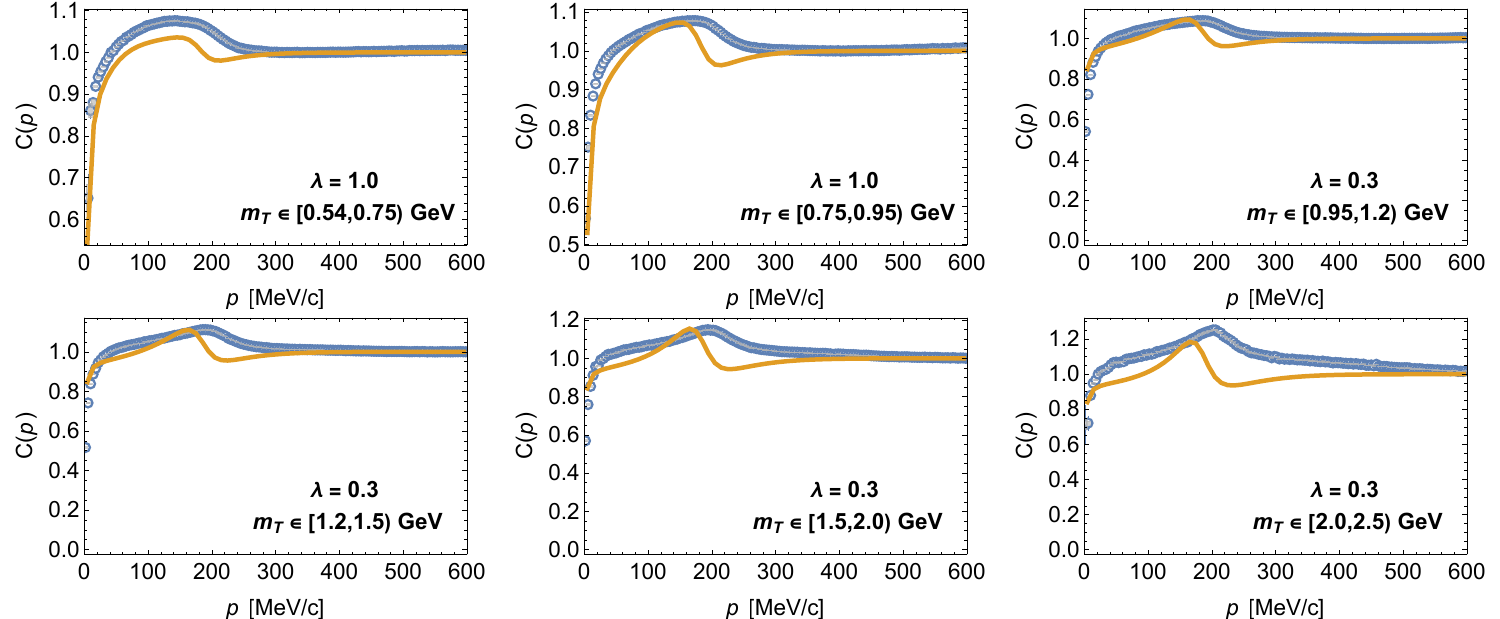}
    \caption{Comparison of ALICE experimental data (blue circles)~\cite{ALICE:2025aur} with the theoretical baselines (yellow lines) derived from the Friedrichs-Lee model and the Koonin-Pratt formula [Eq.~\eqref{eq:KP}]. The calculation employs various Gaussian sources with $R \simeq 0.8\,\text{to}\,2.5$~fm, consistent with the radii extracted from the experimental fit in Ref.~\cite{ALICE:2025aur}. 
    }
    \label{fig:baseline}
\end{figure*}

To validate the quantitative performance of our framework, the corresponding Schr\"odinger equation is solved in coordinate space, including both Coulomb and background interactions~\cite{ALICE:2025aur}. Gaussian sources with $R \simeq 0.8\,\text{to}\,2.5$~fm are employed, consistent with the radii extracted from experimental fits in different $m_T$ regions in Ref.~\cite{ALICE:2025aur}, and the results are compared with experimental data in Fig.~\ref{fig:baseline}. The theoretical calculations capture the widths and positions of the experimental peaks across all $m_T$ regions, but underestimate the peak amplitudes and predict high-momentum side dips absent in the data. To facilitate the comparison, a strength parameter $\lambda$ is introduced via $C_{\text{tuned}}(p)=1+\lambda(C(p)-1)$, which scales the correlation without affecting its shape.

These residual discrepancies point to source properties beyond the simple Gaussian assumption. 
In particular, the predicted high-momentum side dip is a robust consequence of the off-shell dispersive structure [Eq.~\eqref{eq:Cint-offshell}]. Its absence in the data suggests that the actual source imprints compensating properties on the correlation function. 
Resolving this tension calls for a more detailed characterization of the source, which is left for future work.
Nevertheless, establishing such a baseline from the resonance dynamics is a prerequisite for understanding the final-stage evolution of high-energy collisions via femtoscopy. 


\emph{Summary}.{\bf ---}\label{sec:summary}
The ALICE Collaboration has reported a shift of the $\Delta(1232)$ resonance peak in both $\pi^+$-$p$ and $\pi^+$-$d$ femtoscopic correlations~\cite{ALICE:2025aur,ALICE:2025zzg}.
Revisiting resonance contributions from scattering dynamics, we establish a scattering-based baseline in which off-shell effects generate a dispersive structure naturally linked to the observed peak shift.

We formulate the problem by embedding the T-matrix into the Koonin-Pratt framework.
The scattering wave function, constructed via the Lippmann-Schwinger equation, leads to a decomposition of the correlation function into an interference term $C_{\mathrm{int}}$ and a scattering term $C_{\mathrm{scat}}$.
The finite source size induces non-diagonal momentum-space couplings, providing direct sensitivity to off-shell dynamics.
Using the Sokhotski-Plemelj theorem, we separate on-shell and off-shell contributions in a controlled manner.
We find that the on-shell part of $C_{\mathrm{int}}$ exhibits a negative Breit-Wigner structure that largely cancels the full $C_{\mathrm{scat}}$, leaving the off-shell $C_{\mathrm{int}}$ as the dominant contribution.
Its dispersive component, proportional to $M_R(E)-E$, is positive below and negative above the resonance, producing a peak-dip structure across the resonance region.

We compute the $\pi$-$p$ correlation functions within a Friedrichs-Lee framework constrained by $\pi^+$-$p$ scattering data, employing Gaussian emission sources.
The calculation reproduces the observed peak position and width without invoking any in-medium modification of the $\Delta$ resonance, demonstrating that the peak shift can arise from off-shell scattering dynamics alone.
However, the predicted correlation strength remains below the measured amplitude, requiring a phenomenological parameter $\lambda$ to rescale the overall magnitude without altering the spectral shape.
Moreover, the predicted high-momentum dip, as a companion signature of the dispersive structure, is not observed experimentally.

These results establish off-shell dynamics as the minimal baseline for resonance femtoscopy and place nontrivial constraints on interpretations of the $\Delta$ peak shift.
The remaining discrepancies point to missing structure in the emission source that cannot be absorbed into simple normalization effects.
A consistent description therefore calls for a unified treatment of source dynamics and scattering evolution.

\vspace{1.cm}
\emph{Acknowledgments}.{\bf ---}
This work was supported in part by the National Natural Science Foundation of China under contract Nos. 12547102,  12275054, 12147101, 12061141008, 12347106, 12422509, 12375121, and 12547129, and National Key R\&D Program of China under Grant No. 2024YFA1610802 and 2018YFE0104600, Shanghai Pilot Program for Basic Research - Fudan University 21TQ1400100(22TQ006) and the STCSM under Grant No. 23590780100 and ~23JC1400200.

\vspace{1.cm}
\end{CJK*}
\bibliography{04}

@article{XiBS:NST,
  title = {Study of the momentum correlation of nucleons in {{$^{96}_{44}$Ru + $^{96}_{44}$Ru and $^{96}_{40}$Zr + $^{96}_{40}$Zr}} collisions at {{$\sqrt{s_{NN}}=7.7$ and 200 GeV}} from a multiphase transport model},
  author = {Xi, Bao-Shan and Chen, Jin-Hui and Ma, Long and Ma, Yu-Gang and Wang, Ting-Ting},
  year = 2025,
  journal = {Nucl. Sci. Tech.},
  volume = {36},
  pages = {228},
  doi = {10.1007/s41365-025-01826-w}
  }

@article{STAR:dL,
  title = {First Observation of Deuteron-{{$\Lambda$}} Correlations at RHIC},
  author = {B. E. Aboona and  J. Adam and L. Adamczyk and others},
  year = 2026,
      collaboration = "STAR",
  journal = {Phys. Rev. Lett.},
  volume = {136},
  pages = {242303},
  doi = {10.1103/3m26-5y83}
  }

@article{WangTT:PRC24,
  title = {Calculation of momentum correlation functions between $\pi$, {{$K$}}, and $p$ for several heavy-ion collision systems at {{$\sqrt{s_{NN}}=39$ GeV}}},
  author = {Wang, Ting-Ting and Ma, Yu-Gang and  Zhang, S. },
  year = 2024,
  journal = {Phys. Rev. C},
  volume = {109},
  pages = {024912},
  doi = {10.1103/PhysRevC.109.024912}
  }

@article{WangTT:PRC23,
  title = {Simulations of momentum correlation functions of light (anti)nuclei in relativistic heavy-ion collisions at {{$\sqrt{s_{NN}}=39$ GeV}}},
  author = {Wang, Ting-Ting and Ma, Yu-Gang and  Zhang, S. },
  year = 2023,
  journal = {Phys. Rev. C},
  volume = {107},
  pages = {014911},
  doi = {10.1103/PhysRevC.107.014911}
  }

@article{MaYG:PRC06,
  title = {Surveying the nucleon-nucleon momentum correlation function in the framework of quantum molecular dynamics model },
  author = {Ma, Y. G. and  Wei, Y. B. and Shen, W. Q.  and others },
  year = 2006,
  journal = {Phys. Rev. C},
  volume = {73},
  pages = {014604},
  doi = {10.1103/PhysRevC.73.014604}
  }

@article{DingMQ:NST,
  title = {Neutron skin and its effects in heavy-ion collisions},
  author = {Ding, M. Q. and  Fang, D. Q. and  Ma, Y. G. },
  year = 2024,
  journal = {Nucl. Sci. Tech.},
  volume = {35},
  pages = {211},
  doi = {10.1007/s41365-024-01584-1}
  }

@article{FangDQ:PRC16,
  title = {Proton-proton correlations in distinguishing the two-proton emission mechanism of {{$^{23}$Al and $^{22}$Mg}} },
  author = {Fang, D. Q. and Ma, Y. G. and Sun, X. Y. and others },
  year = 2016,
  journal = {Phys. Rev. C},
  volume = {94},
  pages = {044621},
  doi = {10.1103/PhysRevC.94.044621}
  }

@article{Fabbietti:2020bfg,
    author = "Fabbietti, L. and Mantovani Sarti, V. and Vazquez Doce, O.",
    title = "{Study of the Strong Interaction Among Hadrons with Correlations at the LHC}",
    eprint = "2012.09806",
    archivePrefix = "arXiv",
    primaryClass = "nucl-ex",
    doi = "10.1146/annurev-nucl-102419-034438",
    journal = "Ann. Rev. Nucl. Part. Sci.",
    volume = "71",
    pages = "377--402",
    year = "2021"
}

@article{STAR:2015kha,
  title = {Measurement of Interaction between Antiprotons},
  year = 2015,
  journal = {Nature},
  volume = {527},
  number = {7578},
  pages = {345},
  issn = {0028-0836, 1476-4687},
  doi = {10.1038/nature15724},
  urldate = {2025-06-16},
  langid = {english},
  author = {{STAR Collaboration}}
}

@article{Lisa:2005dd,
  title = {{{FEMTOSCOPY IN RELATIVISTIC HEAVY ION COLLISIONS}}: Two Decades of Progress},
  shorttitle = {Femtoscopy in Relativistic Heavy Ion Collisions},
  author = {Lisa, Michael Annan and Pratt, Scott and Soltz, Ron and Wiedemann, Urs},
  year = 2005,
  journal = {Ann. Rev. Nucl. Part. Sci.},
  volume = {55},
  number = {1},
  pages = {357--402},
  issn = {0163-8998, 1545-4134},
  doi = {10.1146/annurev.nucl.55.090704.151533},
  urldate = {2025-06-16},
  langid = {english}
}

@article{vanHees:2004vt,
    author = "van Hees, Hendrik and Rapp, Ralf",
    title = "{Delta(1232) and nucleon spectral functions in hot hadronic matter}",
    eprint = "nucl-th/0407050",
    archivePrefix = "arXiv",
    doi = "10.1016/j.physletb.2004.10.062",
    journal = "Phys. Lett. B",
    volume = "606",
    pages = "59--66",
    year = "2005"
}

@article{Adamczewski-Musch:2020edy,
  title = {Correlated Pion-Proton Pair Emission off Hot and Dense {{QCD}} Matter},
  author = {{Adamczewski-Musch}, J. and Arnold, O. and Behnke, C. and others},
  year = 2021,
  month = aug,
  journal = {Phys. Lett. B},
  volume = {819},
  primaryclass = {nucl-ex},
  pages = {136421},
  issn = {0370-2693},
  doi = {10.1016/j.physletb.2021.136421},
  urldate = {2025-12-28},
  langid = {english}
}

@article{ALICE:2019gcn,
  title = {Scattering Studies with Low-Energy Kaon-Proton Femtoscopy in Proton-Proton Collisions at the {{LHC}}},
  year = 2020,
  month = mar,
  journal = {Phys. Rev. Lett.},
  volume = {124},
  number = {9},
  eprintclass = {nucl-ex},
  pages = {92301},
  publisher = {American Physical Society},
  issn = {0031-9007, 1079-7114},
  doi = {10.1103/PhysRevLett.124.092301},
  urldate = {2025-11-18},
  langid = {english},
  author = {{ALICE Collaboration}}
}

@article{ALICE:2020mfd,
  title = {Unveiling the Strong Interaction among Hadrons at the {{LHC}}},
  year = 2021,
  month = jan,
  journal = {Nature},
  volume = {588},
  number = {7837},
  eprintclass = {nucl-ex},
  pages = {232--238},
  publisher = {Nature Research},
  issn = {0028-0836, 1476-4687},
  doi = {10.1038/s41586-020-3001-6},
  urldate = {2025-06-16},
  langid = {english},
  pmid = {33299194},
  author = {{ALICE Collaboration}}
}

@article{ALICE:2021cpv,
  title = {Experimental Evidence for an Attractive $p$ - {$\phi$} Interaction},
  year = 2021,
  month = oct,
  journal = {Phys. Rev. Lett.},
  volume = {127},
  number = {17},
  eprintclass = {nucl-ex},
  pages = {172301},
  issn = {0031-9007, 1079-7114},
  doi = {10.1103/PhysRevLett.127.172301},
  urldate = {2025-06-16},
  langid = {english},
  author = {{ALICE Collaboration}}
}

@article{ALICE:2023sjd,
  title = {Common Femtoscopic Hadron-Emission Source in $pp$ Collisions at the {{LHC}}},
  year = 2025,
  month = feb,
  journal = {Eur. Phys. J. C},
  volume = {85},
  number = {2},
  eprintclass = {hep-ph},
  pages = {198},
  issn = {1434-6052},
  doi = {10.1140/epjc/s10052-025-13793-y},
  urldate = {2025-06-16},
  langid = {english},
  author = {{ALICE Collaboration}}
}

@article{ALICE:2025aur,
  title = {Investigating the  $p-\pi^{\pm}$ 
 and $p-p-\pi^{\pm}$ dynamics with Femtoscopy in $pp$ Collisions at {{$\sqrt{s}$ = 13 TeV}}},
  year = 2025,
  month = aug,
  journal = {Eur. Phys. J. A},
  volume = {61},
  number = {8},
  eprintclass = {nucl-ex},
  pages = {194},
  issn = {1434-601X},
  doi = {10.1140/epja/s10050-025-01615-4},
  urldate = {2025-12-24},
  langid = {english},
  author = {{ALICE Collaboration}}
}

@article{ALICE:2025byl,
  title = {Observation of Deuteron and Antideuteron Formation from Resonance-Decay Nucleons},
  year = 2025,
  month = dec,
  journal = {Nature},
  volume = {648},
  number = {8093},
  eprintclass = {nucl-ex},
  pages = {306--311},
  publisher = {Nature Publishing Group},
  issn = {1476-4687},
  doi = {10.1038/s41586-025-09775-5},
  urldate = {2025-12-24},
  copyright = {2025 The Author(s)},
  langid = {english},
  author = {{ALICE Collaboration}}
}

@article{ALICE:2025zzg,
  title = {Accessing the Deuteron Source with Pion-Deuteron Femtoscopy in {{Pb-Pb}} Collisions at {{$\sqrt{s_{NN}} =5.02$ TeV}}},
  year = 2025,
  month = dec,
  journal = {Phys. Rev. C},
  volume = {112},
  number = {6},
  primaryclass = {nucl-ex},
  pages = {64003},
  publisher = {American Physical Society},
  doi = {10.1103/mrp4-z4hh},
  urldate = {2026-04-29},
  author = {{ALICE Collaboration}}
}

@article{Breit:1936zzb,
  title = {Capture of {{Slow Neutrons}}},
  author = {Breit, G. and Wigner, E.},
  year = 1936,
  journal = {Phys. Rev.},
  volume = {49},
  number = {7},
  pages = {519--531},
  publisher = {American Physical Society},
  doi = {10.1103/PhysRev.49.519},
  urldate = {2026-03-12},
  langid = {american}
}

@article{Carter:1971tj,
  title = {The Total Cross Sections for Pion-Proton Scattering between 70 {{MeV}} and 290 {{MeV}}},
  author = {Carter, A. A. and Williams, J. R. and Bugg, D. V. and Bussey, P. J. and Dance, D. R.},
  year = 1971,
  journal = {Nucl. Phys. B},
  volume = {26},
  number = {3},
  pages = {445--460},
  issn = {0550-3213},
  doi = {10.1016/0550-3213(71)90188-X},
  urldate = {2026-03-23},
  langid = {american}
}

@article{Chapman:1995nz,
  title = {Extracting Source Parameters from Gaussian Fits to Two-Particle Correlations},
  author = {Chapman, Scott and Nix, J. Rayford and Heinz, Ulrich},
  year = 1995,
  month = nov,
  journal = {Phys. Rev. C},
  volume = {52},
  number = {5},
  pages = {2694--2703},
  issn = {0556-2813, 1089-490X},
  doi = {10.1103/PhysRevC.52.2694},
  urldate = {2025-06-16},
  copyright = {http://link.aps.org/licenses/aps-default-license},
  langid = {english}
}

@article{Dominguez:1980jv,
  title = {{{$\pi NN$, $\pi N\Delta$ and $\pi\pi\rho$}} Form Factors from Quasi-Two-Body Hadronic Reactions},
  author = {Dominguez, C. A.},
  year = 1981,
  journal = {Phys. Rev. C},
  volume = {24},
  number = {6},
  pages = {2611--2617},
  publisher = {American Physical Society},
  doi = {10.1103/PhysRevC.24.2611},
  urldate = {2026-03-10}
}

@article{Epelbaum:2025aan,
  title = {Can the Strong Interactions between Hadrons Be Determined Using Femtoscopy?},
  author = {Epelbaum, Evgeny and Heihoff, Sven and Mei{\ss}ner, Ulf-G. and Tscherwon, Alexander},
  year = 2026,
  month = may,
  journal = {Phys. Rev. Lett.},
  volume = {136},
  number = {21},
  pages = {212301},
  publisher = {American Physical Society},
  doi = {10.1103/tfsb-wlsd},
  urldate = {2026-06-05},
  langid = {english}
}

@article{friedrichs_perturbation_1948,
  title = {On the Perturbation of Continuous Spectra},
  author = {Friedrichs, K. O.},
  year = 1948,
  journal = {Commun. Pure Appl. Math.},
  volume = {1},
  number = {4},
  pages = {361--406},
  issn = {1097-0312},
  doi = {10.1002/cpa.3160010404},
  urldate = {2026-04-22},
  copyright = {Copyright \copyright{} 1948 Wiley Periodicals, Inc., A Wiley Company},
  langid = {english}
}

@article{Giacosa:2021mbz,
  title = {A Simple Alternative to the Relativistic {{Breit}}--{{Wigner}} Distribution},
  author = {Giacosa, Francesco and Okopi{\'n}ska, Anna and Shastry, Vanamali},
  year = 2021,
  month = dec,
  journal = {Eur. Phys. J. A},
  volume = {57},
  number = {12},
  primaryclass = {hep-ph},
  pages = {336},
  issn = {1434-601X},
  doi = {10.1140/epja/s10050-021-00641-2},
  urldate = {2026-03-18},
  langid = {english}
}

@article{Goldhaber:1960sf,
  title = {Influence of {Bose-Einstein} Statistics on the Antiproton-Proton Annihilation Process},
  author = {Goldhaber, Gerson and Goldhaber, Sulamith and Lee, Wonyong and Pais, Abraham},
  year = 1960,
  month = oct,
  journal = {Phys. Rev.},
  volume = {120},
  number = {1},
  pages = {300--312},
  publisher = {American Physical Society},
  issn = {0031-899X},
  doi = {10.1103/PhysRev.120.300},
  urldate = {2025-06-16},
  copyright = {http://link.aps.org/licenses/aps-default-license},
  langid = {english}
}

@article{Gridnev:2018rbw,
  title = {Charge Splitting in {{$\pi$N$\Delta$}} Form Factor},
  author = {Gridnev, A. B. and Abaev, V. V. and Kozlenko, N. G.},
  year = 2019,
  month = apr,
  journal = {Phys. Atom. Nucl.},
  volume = {82},
  number = {2},
  primaryclass = {hep-ph},
  pages = {163--167},
  doi = {10.1134/S1063778819020091},
  urldate = {2026-03-10},
  langid = {english}
}

@article{HanburyBrown:1956bqd,
  title = {A Test of a New Type of Stellar Interferometer on Sirius},
  author = {Hanbury Brown, R. and Twiss, R. Q.},
  year = 1956,
  journal = {Nature},
  volume = {178},
  number = {4541},
  pages = {1046--1048},
  publisher = {Nature Publishing Group},
  issn = {0028-0836, 1476-4687},
  doi = {10.1038/1781046a0},
  urldate = {2025-06-16},
  copyright = {http://www.springer.com/tdm},
  langid = {english}
}

@article{Haskins:1985xg,
  title = {Breit--{{Wigner}} Resonance and the Delta++},
  author = {Haskins, J. Richard},
  year = 1985,
  month = oct,
  journal = {Am. J. Phys.},
  volume = {53},
  number = {10},
  pages = {988--991},
  issn = {0002-9505},
  doi = {10.1119/1.14017},
  urldate = {2026-03-23},
  langid = {american}
}

@article{Hoferichter:2023mgya,
  title = {Nucleon Resonance Parameters from {Roy--Steiner} Equations},
  author = {Hoferichter, Martin and Ruiz De Elvira, Jacobo and Kubis, Bastian and Mei{\ss}ner, Ulf-G.},
  year = 2024,
  month = may,
  journal = {Phys. Lett. B},
  volume = {853},
  primaryclass = {hep-ph},
  pages = {138698},
  issn = {03702693},
  doi = {10.1016/j.physletb.2024.138698},
  urldate = {2026-03-24},
  langid = {english}
}

@article{Hunt:2018wqz,
  title = {Updated Determination of ${{N}}^*$ Resonance Parameters Using a Unitary, Multichannel Formalism},
  author = {Hunt, B. C. and Manley, D. M.},
  year = 2019,
  month = may,
  journal = {Phys. Rev. C},
  volume = {99},
  number = {5},
  primaryclass = {nucl-ex},
  pages = {55205},
  publisher = {American Physical Society},
  doi = {10.1103/PhysRevC.99.055205},
  urldate = {2026-03-24}
}

@article{Kamiya:2019uiw,
  title = {{{$K$-$p$}} Correlation Function from High-Energy Nuclear Collisions and Chiral {{SU}}(3) Dynamics},
  author = {Kamiya, Yuki and Hyodo, Tetsuo and Morita, Kenji and Ohnishi, Akira and Weise, Wolfram},
  year = 2020,
  month = apr,
  journal = {Phys. Rev. Lett.},
  volume = {124},
  number = {13},
  eprintclass = {nucl-th},
  pages = {132501},
  publisher = {American Physical Society},
  issn = {0031-9007, 1079-7114},
  doi = {10.1103/PhysRevLett.124.132501},
  urldate = {2025-06-16},
  langid = {english},
  pmid = {32302166}
}

@article{Koonin:1977fh,
  title = {Proton Pictures of High-Energy Nuclear Collisions},
  author = {Koonin, Steven E.},
  year = 1977,
  journal = {Phys. Lett. B},
  volume = {70},
  number = {1},
  pages = {43--47},
  issn = {03702693},
  doi = {10.1016/0370-2693(77)90340-9},
  urldate = {2025-06-16},
  copyright = {https://www.elsevier.com/tdm/userlicense/1.0/},
  langid = {english}
}

@article{Kopylov:1974th,
  title = {Like Particle Correlations as a Tool to Study the Multiple Production Mechanism},
  author = {Kopylov, G.I.},
  year = 1974,
  month = jun,
  journal = {Phys. Lett. B},
  volume = {50},
  number = {4},
  pages = {472--474},
  issn = {03702693},
  doi = {10.1016/0370-2693(74)90263-9},
  urldate = {2025-06-16},
  copyright = {https://www.elsevier.com/tdm/userlicense/1.0/},
  langid = {english}
}

@inproceedings{lednicky_femtoscopy_2001,
  title = {Femtoscopy with Unlike Particles},
  booktitle = {Int. {{Workshop Phys}}. {{Quark Gluon Plasma}}},
  author = {Lednicky, R.},
  year = 2001,
  month = dec,
  langid = {english}
}

@article{lednicky_influence_1996,
  title = {The Influence of Final State Interaction on Two-Particle Correlations in Multiple Production of Particles and Resonances},
  author = {Lednick{\'y}, R. and Lyuboshitz, V. L.},
  year = 1996,
  month = apr,
  journal = {Acta Phys. Hung. New Ser. Heavy Ion Phys.},
  volume = {3},
  number = {1},
  pages = {93--113},
  issn = {1588-2675},
  doi = {10.1007/BF03053635},
  urldate = {2025-06-16},
  copyright = {https://www.springer.com/tdm},
  langid = {english}
}

@inproceedings{Lednicky:2002fq,
  title = {Progress in Correlation Femtoscopy},
  booktitle = {Multiparticle {{Dyn}}.},
  author = {Lednicky, R.},
  year = 2003,
  month = jul,
  pages = {21--26},
  publisher = {WORLD SCIENTIFIC},
  address = {Alushta, Crimea, Ukraine},
  doi = {10.1142/9789812704962_0005},
  urldate = {2025-06-16},
  isbn = {978-981-238-403-4 978-981-270-496-2},
  langid = {english}
}

@article{Lee:1954iq,
  title = {Some Special Examples in Renormalizable Field Theory},
  author = {Lee, T. D.},
  year = 1954,
  month = sep,
  journal = {Phys. Rev.},
  volume = {95},
  number = {5},
  pages = {1329--1334},
  publisher = {American Physical Society},
  doi = {10.1103/PhysRev.95.1329},
  urldate = {2026-04-22}
}

@article{ParticleDataGroup:2024cfk,
  title = {Review of Particle Physics},
  year = 2024,
  month = aug,
  journal = {Phys. Rev. D},
  volume = {110},
  number = {3},
  pages = {030001},
  publisher = {American Physical Society},
  doi = {10.1103/PhysRevD.110.030001},
  urldate = {2024-11-12},
  langid = {english},
  author = {{Particle Data Group Collaboration}}
}

@article{Pratt:1986cc,
  title = {Pion Interferometry of Quark-Gluon Plasma},
  author = {Pratt, Scott},
  year = 1986,
  journal = {Phys. Rev. D: Part. Fields},
  volume = {33},
  number = {5},
  pages = {1314--1327},
  publisher = {American Physical Society},
  issn = {0556-2821},
  doi = {10.1103/PhysRevD.33.1314},
  urldate = {2025-06-16},
  copyright = {http://link.aps.org/licenses/aps-default-license},
  langid = {english}
}

@article{Pratt:1986ev,
  title = {Coherence and Coulomb Effects on Pion Interferometry},
  author = {Pratt, Scott},
  year = 1986,
  month = jan,
  journal = {Phys. Rev. D},
  volume = {33},
  number = {1},
  pages = {72--79},
  issn = {0556-2821},
  doi = {10.1103/PhysRevD.33.72},
  urldate = {2025-06-16},
  copyright = {http://link.aps.org/licenses/aps-default-license},
  langid = {english}
}

@article{Reichert:2019lny,
  title = {Delta Mass Shift as a Thermometer of Kinetic Decoupling in {{Au}} + {{Au}} Reactions at 1.23 {{AGeV}}},
  author = {Reichert, Tom and Hillmann, Paula and Limphirat, Ayut and Herold, Christoph and Bleicher, Marcus},
  year = 2019,
  month = aug,
  journal = {J. Phys. G},
  volume = {46},
  number = {10},
  primaryclass = {nucl-th},
  pages = {105107},
  publisher = {IOP Publishing},
  issn = {0954-3899},
  doi = {10.1088/1361-6471/ab34fa},
  urldate = {2025-12-28},
  langid = {english}
}

@article{Wiedemann:1999qn,
  title = {Particle Interferometry for Relativistic Heavy-Ion Collisions},
  author = {Wiedemann, Urs Achim and Heinz, Ulrich},
  year = 1999,
  month = oct,
  journal = {Phys. Rep.},
  volume = {319},
  number = {4-5},
  pages = {145--230},
  issn = {03701573},
  doi = {10.1016/S0370-1573(99)00032-0},
  urldate = {2025-06-16},
  langid = {english}
}

@article{Xiao:2023lpv,
  title = {On the Generalized {{Friedrichs-Lee}} Model with Multiple Discrete and Continuous States},
  author = {Xiao, Zhiguang and Zhou, Zhi-Yong},
  year = 2025,
  month = aug,
  journal = {Chin. Phys. C},
  volume = {49},
  number = {8},
  pages = {83102},
  issn = {1674-1137, 2058-6132},
  doi = {10.1088/1674-1137/adcd4b},
  urldate = {2026-04-22},
  langid = {english}
}

@misc{Xie:2026hpp,
  title = {Off-Shell Chiral Dynamics in the {{$\Lambda$}}(1405) Resonance and ${{K}}^-p$ Femtoscopic Correlations},
  author = {Xie, Jia-Ming and Liu, Zhi-Wei and Lu, Jun-Xu and Liang, Haozhao and Geng, Li-Sheng},
  year = 2026,
  month = apr,
  number = {arXiv:2604.00791},
  eprint = {2604.00791},
  primaryclass = {nucl-th},
  publisher = {arXiv},
  doi = {10.48550/arXiv.2604.00791},
  urldate = {2026-05-20},
  archiveprefix = {arXiv},
  langid = {english}
}

@article{Xu:2015jxa,
  title = {Temperature Dependence of Decuplet Baryon Masses from Thermal {{QCD}} Sum Rules},
  author = {Xu, Yong-Jiang and Liu, Yong-Lu and Huang, Ming-Qiu},
  year = 2015,
  month = feb,
  journal = {Commun. Theor. Phys.},
  volume = {63},
  number = {2},
  primaryclass = {hep-ph},
  pages = {209--214},
  issn = {0253-6102},
  doi = {10.1088/0253-6102/63/2/13},
  urldate = {2026-01-29},
  langid = {english}
}

@article{Yamaguchi:1954mp,
  title = {Two-{{Nucleon Problem When}} the {{Potential Is Nonlocal}} but {{Separable}}. {{I}}},
  author = {Yamaguchi, Yoshio},
  year = 1954,
  journal = {Phys. Rev.},
  volume = {95},
  number = {6},
  pages = {1628--1634},
  doi = {10.1103/PhysRev.95.1628},
  langid = {american}
}

@article{Yamaguchi:1954zz,
  title = {Two-{{Nucleon Problem When}} the {{Potential Is Nonlocal}} but {{Separable}}. {{II}}},
  author = {Yamaguchi, Yoshio and Yamaguchi, Yoriko},
  year = 1954,
  journal = {Phys. Rev.},
  volume = {95},
  number = {6},
  pages = {1635--1643},
  publisher = {American Physical Society},
  doi = {10.1103/PhysRev.95.1635},
  urldate = {2026-01-30}
}

@article{Zajc:1984vb,
  title = {Two-Pion Correlations in Heavy Ion Collisions},
  author = {Zajc, W. A. and Bistirlich, J. A. and Bossingham, R. R. and Bowman, H. R. and Clawson, C. W. and Crowe, K. M. and Frankel, K. A. and Ingersoll, J. G. and Kurck, J. M. and Martoff, C. J. and Murphy, D. L. and Rasmussen, J. O. and Sullivan, J. P. and Yoo, E. and Hashimoto, O. and Koike, M. and McDonald, W. J. and Miller, J. P. and Tru{\"o}l, P.},
  year = 1984,
  month = jun,
  journal = {Phys. Rev. C},
  volume = {29},
  number = {6},
  pages = {2173--2187},
  publisher = {American Physical Society},
  issn = {0556-2813},
  doi = {10.1103/PhysRevC.29.2173},
  urldate = {2026-06-05},
  copyright = {http://link.aps.org/licenses/aps-default-license},
  langid = {english}
}

@misc{Zhang:2025tfd,
  title = {Shedding Light on (Anti-)Nuclei Production with Pion-Nucleus Femtoscopy},
  author = {Zhang, Li-Yuan and Ko, Che Ming and Ma, Yu-Gang and Shou, Qi-Ye and Sun, Kai-Jia and Wang, Rui and Zhang, Song},
  year = 2025,
  month = nov,
  eprint = {2511.10298},
  primaryclass = {nucl-th},
  doi = {10.48550/arXiv.2511.10298},
  urldate = {2026-01-29},
  archiveprefix = {arXiv},
  langid = {english}
}
\end{document}


\title{Supplemental Material}



\maketitle

\section{From the source-incorporated Liouville equation to the Koonin-Pratt formula}
\label{sec:Liouville-to-KP}

The particle pair production in a high-energy nucleus-nucleus collision is described by the quantum Liouville equation for the density matrix $\rho(\tau)$,
\begin{equation}
 \label{eq:liouville}
 \partial_t \rho(\tau) + i[H,\rho(\tau)] = I(\tau).
\end{equation}
Since femtoscopy concerns two-particle correlations, we reduce the system to a two-particle density-matrix description, where $I(\tau)$ acts as a pair-pumping source: particle pairs are continuously produced and subsequently freeze out~\cite{Wiedemann:1999qn}.
With the time-evolution propagator $\mathcal{U}(t,\tau)=e^{iH(t-\tau)}$ and noting the initial state condition, $\rho(-\infty)=0$, the final-state solution reads
\begin{equation}
 \rho(\infty) = \int_{-\infty}^{\infty} d\tau \, \mathcal{U}(-\infty,\tau) I(\tau) \mathcal{U}^\dagger(-\infty,\tau).
\end{equation}

For simplicity, we adopt a minimal model with continuum states $|\vct{k}_1,\vct{k}_2\rangle$ and a single resonance state $|R,\vct{K}\rangle$, the latter representing a short-lived resonance that decays before freeze-out. The source operator then expands as
\begin{equation}
\label{eq:I_expand}
\begin{aligned}
I(\tau) &= \int d^3\vct{k}_1 d^3\vct{k}_2 d^3\vct{k}'_1 d^3\vct{k}'_2 \; |\vct{k}_1,\vct{k}_2\rangle\langle\vct{k}'_1,\vct{k}'_2| \; I_{\text{scat}}(\vct{k}_1,\vct{k}_2;\vct{k}'_1,\vct{k}'_2;\tau) \\
&\quad + \int d^3\vct{K} d^3\vct{K}' \; |R,\vct{K}\rangle\langle R,\vct{K}'| \; I_{\text{res}}(\vct{K};\vct{K}';\tau) \\
&\quad + \int d^3\vct{k}_1 d^3\vct{k}_2 d^3\vct{K} \; \Bigl( |\vct{k}_1,\vct{k}_2\rangle\langle R,\vct{K}| \; I_{\text{res-scat}}(\vct{k}_1,\vct{k}_2;\vct{K};\tau) + \text{h.c.} \Bigr),
\end{aligned}
\end{equation}
representing direct pair production, bare resonance production with subsequent decay, and the coherent Fano interference channel, respectively.

The femtoscopic correlation~\cite{Wiedemann:1999qn,Lisa:2005dd} function is defined as
\begin{equation}
 \label{eq:corr-def}
 C(\vct{p}_1,\vct{p}_2)=\frac{P_2(\vct{p}_1,\vct{p}_2)}{P_1(\vct{p}_1)P_1(\vct{p}_2)} = \frac{\text{Tr}[|\vct{p}_1,\vct{p}_2\rangle\langle \vct{p}_1,\vct{p}_2|\rho(\infty)]}{P_1(\vct{p}_1)P_1(\vct{p}_2)},
\end{equation}
with $P_1$ the single-particle distribution. Substituting Eqs.~\eqref{eq:liouville}--\eqref{eq:I_expand} and defining the time-integrated source $S = \int d\tau\, I(\tau)$, the pair probability $P_2$ inherits the same structure.

The matrix element $\langle \vct{p}_1,\vct{p}_2|\mathcal{U}(\infty,\tau)|\vct{k}_1,\vct{k}_2\rangle$ describes the evolution of a pair from production time $\tau$ to the asymptotic state and is given by the scattering wave function with outgoing boundary conditions:
\begin{equation}
\label{eq:U_to_psi}
\langle \vct{p}_1,\vct{p}_2|\mathcal{U}(\infty,\tau)|\vct{k}_1,\vct{k}_2\rangle
= e^{iE(\infty-\tau)} \, \left(\psi^{(+)}_{\vct{p}_1,\vct{p}_2}(\vct{k}_1,\vct{k}_2)\right)^\dagger.
\end{equation}
The resonance-to-continuum transition follows:
\begin{equation}
\label{eq:U_to_T}
\langle \vct{p}_1,\vct{p}_2|\mathcal{U}(\infty,\tau)|R,\vct{K}\rangle
= e^{iE(\infty-\tau)} \, (T_{R,\vct{K};\vct{p}_1,\vct{p}_2})^\dagger \, (G^{(+)}_R)^\dagger,
\end{equation}
where $T_{R,\vct{K};\vct{p}_1,\vct{p}_2}$ is the resonance–continuum matrix element of the $T$ matrix and $G^{(+)}_R$ denotes the bare resonance propagator.

Substituting Eqs.~\eqref{eq:U_to_psi} and \eqref{eq:U_to_T} yields the Koonin-Pratt formula:
\begin{equation}
\label{eq:P2_full}
\begin{aligned}
P_2(\vct{p}_1,\vct{p}_2) &=
\int d^3\vct{k}_1 d^3\vct{k}_2 d^3\vct{k}'_1 d^3\vct{k}'_2 \; \left(\psi^{(+)}_{\vct{p}_1,\vct{p}_2}(\vct{k}_1,\vct{k}_2)\right)^\dagger S_{\text{scat}}(\vct{k}_1,\vct{k}_2;\vct{k}'_1,\vct{k}'_2) \, \psi^{(+)}_{\vct{p}_1,\vct{p}_2}(\vct{k}'_1,\vct{k}'_2) \\
&\quad + \int d^3\vct{K} d^3\vct{K}' \; T_{R;\vct{p}_1,\vct{p}_2}^\dagger \, (G^{(+)}_R)^\dagger S_{\text{res}}(\vct{K};\vct{K}') \, G^{(+)}_R T_{R;\vct{p}_1,\vct{p}_2} \\
&\quad + \int d^3\vct{k}_1 d^3\vct{k}_2 d^3\vct{K} \; \Bigl[ \left(\psi^{(+)}_{\vct{p}_1,\vct{p}_2}(\vct{k}_1,\vct{k}_2)\right)^\dagger S_{\text{res-scat}}(\vct{k}_1,\vct{k}_2;\vct{K}) \, G^{(+)}_R T_{R;\vct{p}_1,\vct{p}_2} + \text{h.c.} \Bigr].
\end{aligned}
\end{equation}

In this work, we focus on the $\pi$-$p$ continuum channel and neglect the resonance and interference terms. The resonance contribution is suppressed since the short-lived $\Delta$ resonance decays before freeze-out, while the interference term vanishes for a decoherent source. Under this approximation, $P_2$ reduces to the first line of Eq.~\eqref{eq:P2_full}, giving Eq.~\eqref{eq:KP} in the main text. The interference term may give rise to Fano interference in femtoscopy, which we leave for future investigation.


\section{Lippmann-Schwinger equation and Sokhotski-Plemelj decomposition in Koonin-Pratt formula}
\label{sec:LS-SP}

To analyze the correlation in momentum space, we substitute the Lippmann-Schwinger solution for the scattering wave function into the Koonin-Pratt formula.
In the center-of-mass (c.m.) frame, the Lippmann-Schwinger equation reads
\begin{equation}
\label{eq:LS_wave}
\psi^{(+)}_{\vct{p}}(\vct{k})
= \delta^{(3)}(\vct{k} - \vct{p})
+ G^{(+)}(\vct{k}; E_{p}) \, T(\vct{k}, \vct{p}; E_{p}),
\end{equation}
where $\vct{k}$ and $\vct{p}$ are the relative momenta,  with $\vct{p}$ the asymptotic momentum of the scattering state, and $(+)$ denotes outgoing-wave boundary conditions.
The free propagator is
\begin{equation}
\label{eq:G0}
G^{(+)}(\vct{k}; E_{p}) = \frac{1}{E_{p} - E_{k} + i\epsilon},
\end{equation}
with $E_{k} = m_1 + m_2 + k^{*2}/2\mu$ the total c.m.\ energy under non-relativistic approximation and $\mu = m_1 m_2/(m_1+m_2)$ the reduced mass.
We adopt a generic separable $T$-matrix for a single resonance,
\begin{equation}
\label{eq:T_separable}
T(\vct{k}, \vct{p}; E_{p})
= \frac{g(\vct{k}) \, g^*(\vct{p})}
{E_{p} - M_{R0} - \Sigma_0(E_{p}) + i\epsilon},
\end{equation}
where $M_{R0}$ is the bare resonance mass, $\Sigma_0$ the self-energy, and $g(\vct{k})$ the vertex form factor.

Substituting Eq.~\eqref{eq:LS_wave} into the first line of Eq.~\eqref{eq:P2_full}, which describes the direct-pair contribution, the correlation function decomposes as
\begin{equation}
\label{eq:C_decomposed}
C(p) = 1 + C_{\text{int}}(p) + C_{\text{scat}}(p),
\end{equation}
with
\begin{equation}
\label{eq:C_int}
C_{\text{int}}(\vct{p}) = 2\,\operatorname{Re}
\int d^{3}\vct{k}\, d^{3}\vct{k}' \;
\delta^{(3)}(\vct{k}-\vct{p}) \,
S(\vct{k},\vct{k}') \,
G^{(+)}(\vct{k}'; E_{p}) \, T(\vct{k}',\vct{p}; E_{p}),
\end{equation}
\begin{equation}
\label{eq:C_scat}
C_{\text{scat}}(\vct{p}) = \int d^{3}\vct{k}\, d^{3}\vct{k}' \;
\bigl(G^{(+)}(\vct{k}; E_{p}) \, T(\vct{k},\vct{p}; E_{p})\bigr)^{\dagger}
\, S(\vct{k},\vct{k}') \,
G^{(+)}(\vct{k}'; E_{p}) \, T(\vct{k}',\vct{p}; E_{p}),
\end{equation}
where $S \equiv S_{\text{scat}}$ and the normalization by single-particle spectra has been omitted. In bra-ket notation these reduce to Eq.~\eqref{eq:LS+KP} in the main text.

The momentum integrals in Eqs.~\eqref{eq:C_int}-\eqref{eq:C_scat} cannot be evaluated analytically.
We apply the Sokhotski-Plemelj theorem to decompose the propagator:
\begin{equation}
\label{eq:SP}
G^{(+)}(\vct{k}; E_{p}) = \frac{1}{E_{k} - E_{p} + i\epsilon}
= \operatorname{P.V.}\frac{1}{E_{k} - E_{p}} - i\pi\,\delta(E_{k} - E_{p}).
\end{equation}
The $\delta$-function (pole) part enforces $k=p$ and yields the {on-shell} contribution; the principal-value (PV) part captures the {off-shell} contribution.
Consequently,
\begin{equation}
\label{eq:on_off_decomp}
\begin{aligned}
C_{\text{int}} &= C_{\text{int}}^{(\text{on-shell})} + C_{\text{int}}^{(\text{off-shell})},\\
C_{\text{scat}} &= C_{\text{scat}}^{(\text{on-shell})} + C_{\text{scat}}^{(\text{off-shell})}+C_{\text{scat}}^{(\text{on-off})}.
\end{aligned}
\end{equation}
Here,
\begin{equation}
 \label{eq:C-sact-on-off}
 C_{\text{scat}}^{(\text{on-off})}\propto i\pi \int d^3\vct{k} d^3\vct{k}' S(\vct{k},\vct{k}')g^*(\vct{k})g(\vct{k}')\left(\delta(E_p-E_k) \operatorname{P.V.}\frac{1}{E_p-E_{k}'}-\delta(E_p-E_{k'})\operatorname{P.V.}\frac{1}{E_p-E_k}\right),
\end{equation}
when source is symmetric in momentum space and assuming there is no open channel other than the resonance channel, this term vanished.

\subsection{On-shell contributions}

Inserting the pole part of Eq.~\eqref{eq:SP} into Eqs.~\eqref{eq:C_int}-\eqref{eq:C_scat} forces the momentum integration on-shell ($k=p$ or $k'=p$).
The separable $T$-matrix~\eqref{eq:T_separable} satisfies the optical theorem for forward on-shell scattering,
\begin{equation}
\label{eq:optical}
-2\,\operatorname{Im} T(\vct{p},\vct{p}; E_{p})
= 2\pi\,\rho(E_{p})\,|T(\vct{p},\vct{p}; E_{p})|^{2},
\end{equation}
which, with the density of states $\rho(E_p) = \int k^2dk \delta(E_p-E_k)=\mu p$ under the non-relativistic approximation, implies
\begin{equation}
\label{eq:ImSigma}
\operatorname{Im} \Sigma_0(E_{p}) = -\,\pi\,\rho(E_{p})\,|g(\vct{p})|^{2}.
\end{equation}
Using this relation, the on-shell contributions evaluate to
\begin{align}
\label{eq:C_int_on}
C_{\text{int}}^{(\text{on-shell})}(\vct{p})
&= -\frac{\Gamma(E_{p})^{2}/2}{(E_{p}-M_{R}(E_p))^{2} + \Gamma(E_{p})^{2}/4}\,
\bigl(S\,\mathcal{P}_{J_{R}}\bigr)(\vct{p},\vct{p}),\\[4pt]
\label{eq:C_scat_on}
C_{\text{scat}}^{(\text{on-shell})}(\vct{p})
&= +\frac{\Gamma(E_{p})^{2}/4}{(E_{p}-M_{R}(E_p))^{2} + \Gamma(E_{p})^{2}/4}\,
\bigl(\mathcal{P}_{J_{R}}^{\dagger}\,S\,\mathcal{P}_{J_{R}}\bigr)(\vct{p},\vct{p}),
\end{align}
where $M_{R}(E_p) = M_{R0} + \operatorname{Re}\Sigma_0(E_{p})$ and $\Gamma(E_{p}) = -2\operatorname{Im}\Sigma_0(E_{p})$ are the physical resonance mass and width, and $\mathcal{P}_{J_{R}}$ projects onto the resonance spin-parity channel. These are Eqs.~\eqref{eq:Cint-on-shell}-\eqref{eq:Cscat-on-shell} in the main text.

Equation~\eqref{eq:C_scat_on} has the standard Breit-Wigner form adopted in conventional femtoscopy~\cite{Kamiya:2019uiw}. However, $C_{\text{int}}^{(\text{on-shell})}$ is {negative} with {twice} the magnitude of $C_{\text{scat}}^{(\text{on-shell})}$, so their sum is a {dip} rather than a peak. This reveals a fundamental incompleteness of the pure on-shell treatment~\cite{lednicky_influence_1996,Kamiya:2019uiw}.

\subsection{Off-shell contributions and net effect}

Substituting the PV part of Eq.~\eqref{eq:SP} yields the off-shell contributions. Although the PV integrals cannot be evaluated in closed form, their general structure is revealing:
\begin{align}
\label{eq:C_int_off}
C_{\text{int}}^{(\text{off-shell})}(\vct{p})
&\propto \frac{M_{R}(E_p)-E_{p}}{(E_{p}-M_{R}(E_p))^{2} + \Gamma(E_{p})^{2}/4}\,
\times \text{P.V.}\int d^3\vct{k} \, (S\mathcal{P}_{J_R})(\vct{p},\vct{k}) \frac{g(\vct{k})g^*(\vct{p})}{E_k-E_p},\\[4pt]
\label{eq:C_scat_off}
C_{\text{scat}}^{(\text{off-shell})}(\vct{p})
&\propto \frac{\Gamma(E_{p})/2\pi\rho(E_{p})}{(E_{p}-M_{R}(E_p))^{2} + \Gamma(E_{p})^{2}/4}\,
\times \text{P.V.}\int d^3\vct{k} d^3\vct{k}' \, (\mathcal{P}^{\dagger}_{J_R} S\mathcal{P}_{J_R})(\vct{k},\vct{k}') \frac{g^*(\vct{k})g(\vct{k}')}{(E_p-E_k)(E_p-E_{k'})}.
\end{align}
In the main text, the PV integrals are absorbed into source-dependent form factors $\mathcal{F}_{\text{int}}$ [Eq.~\eqref{eq:Cint-offshell}] and $\mathcal{F}_{\text{scat}}$ [Eq.~\eqref{eq:Cscat-offshell}] in the main text, which are positive real-valued for the Gaussian source and Friedrichs-Lee model (main text Fig.~\ref{fig:Corr-Apart}).

Two structural features stand out. First, $C_{\text{int}}^{(\text{off-shell})}$ carries a {dispersive} prefactor $(M_{R}(E_p)-E_{p})$, producing an asymmetric peak-dip structure straddling the nominal Breit-Wigner mass. Second, $C_{\text{scat}}^{(\text{off-shell})}$ is a skewed Breit-Wigner-like peak leaning toward lower relative momentum.

Crucially, as shown in the Fig.~\ref{fig:Corr-Apart}, the three subleading terms approximately cancel near the nominal resonance mass:
\begin{equation}
C_{\text{scat}}^{(\text{off-shell})}(\vct{p}_{BW}) + C_{\text{int}}^{(\text{on-shell})}(\vct{p}_{BW}) + C_{\text{scat}}^{(\text{on-shell})}(\vct{p}_{BW}) \simeq 0,
\end{equation}
where $\vct{p}_{BW}$ is the momentum corresponding to the Breit-Wigner mass of resonance (dashed line in the main text Fig.~\ref{fig:Corr-Apart}). The standard on-shell Breit-Wigner peak is thus largely cancelled, and the dispersive interference term $C_{\text{int}}^{(\text{off-shell})}$ emerges as the {leading-order} contribution, imprinting a low-momentum enhancement and a high-momentum depletion. Given the ALICE observation that the $\Delta$ resonance peak in $\pi$-$p$ and $\pi$-$d$ femtoscopy is shifted toward lower relative momentum~\cite{ALICE:2025aur,ALICE:2025byl}, this off-shell dispersive contribution is phenomenologically indispensable.

\section{Friedrichs-Lee model and numerical calculation}
\label{sec:FL-model}

To make the qualitative analysis of Sec.~\ref{sec:LS-SP} quantitative, we implement the Friedrichs-Lee model~\cite{friedrichs_perturbation_1948,Lee:1954iq} for the $\Delta$ resonance in $\pi$-$p$ scattering, combined with a Gaussian source.

In the c.m.\ frame, the Friedrichs-Lee Hamiltonian consists of a free part and a resonance-continuum coupling:
\begin{equation}
\label{eq:FL_Hamilton}
\begin{aligned}
H_0 &= \sum_{s_z} \int d^{3}\vct{k}\; E_{k}\,|\vct{k};s_z\rangle\langle\vct{k};s_z|
      + \sum_{J_z} M_{0}\,|\Delta;J_z\rangle\langle\Delta;J_z|,\\[4pt]
V   &= \sum_{J_z,\,s_z} \int d^{3}\vct{k}\;
      \Bigl[ g_{J_z,s_z}(\vct{k})\,|\Delta;J_z\rangle\langle\vct{k};s_z| + \text{h.c.} \Bigr],
\end{aligned}
\end{equation}
where $s_z=\pm 1/2$ labels the nucleon spin projection, $J_z$ the $\Delta$ spin projection ($J=3/2$), $E_k = m_\pi + m_p + k^{2}/2\mu$, and $M_0$ is the bare $\Delta$ mass.
The vertex function carries the $P$-wave structure of the $\Delta(1232)$ with a dipole form factor:
\begin{equation}
\label{eq:FL_vertex}
\begin{aligned}
g_{J_z,s_z}(\vct{k}) &= g_0\,\frac{k}{m_\pi}\,f(k)\,
Y_{1,\,J_z-s_z}(\hat{\vct{k}})\,
\langle \tfrac{3}{2},J_z | \tfrac{1}{2},s_z; 1, J_z-s_z \rangle,\\[4pt]
f(k) &= \bigl(1 + k^{2}/\Lambda^{2}\bigr)^{-2},
\end{aligned}
\end{equation}
where $g_0$ is the bare coupling, $\Lambda$ the ultraviolet cutoff, and the spherical harmonic $Y_{1,m}$ together with the Clebsch-Gordan coefficient project onto the $P_{33}$ channel.

Solving the Lippmann-Schwinger equation $T = V + V G^{(+)}_0 T$ with this Hamiltonian, and projecting onto the continuum and resonance states, yields the continuum-to-continuum $T$-matrix in separable form:
\begin{align}
\label{eq:T_kk}
T^{s_z,s_z'}_{\vct{k},\vct{k}'}(E)
&\equiv \langle\vct{k};s_z|T(E)|\vct{k}';s_z'\rangle \nonumber\\
&= \sum_{J_z} g_{J_z,s_z}^*(\vct{k})\,
   \frac{1}{E - M_0 - \Sigma_0(E) + i\epsilon}\,
   g_{J_z,s_z'}(\vct{k}') \nonumber\\
&= T_{3/2}(k,k';E)\,
   \mathcal{P}^{s_z,s_z'}_{3/2}(\hat{\vct{k}},\hat{\vct{k}}'),
\end{align}
where the reduced $T$-matrix and the spin projector are
\begin{equation}
\label{eq:T32}
T_{3/2}(k,k';E) = \frac{(g_0/m_\pi)^2\, k\,k'\,f(k)\,f(k')}
{E - M_0 - \Sigma_0(E) + i\epsilon},
\end{equation}
\begin{equation}
\label{eq:P32}
\mathcal{P}^{s_z,s_z'}_{3/2}(\hat{\vct{k}},\hat{\vct{k}}')
= \sum_{J_z} Y_{1,J_z-s_z}^{*}(\hat{\vct{k}}')\,
   Y_{1,J_z-s_z'}(\hat{\vct{k}})\,
   \langle \tfrac{3}{2},J_z | \tfrac{1}{2},s_z; 1, J_z-s_z \rangle\,
   \langle \tfrac{1}{2},s_z'; 1, J_z-s_z' | \tfrac{3}{2},J_z \rangle.
\end{equation}
The self-energy generated by the resonance-continuum coupling is the bubble diagram
\begin{equation}
\label{eq:Sigma0}
\Sigma_0(E) = \int_{0}^{\infty} k'^{2} dk'\;
\frac{\bigl(g_0\,(k'/m_\pi)\,f(k')\bigr)^{2}}
{E - E_{k'} + i\epsilon},
\end{equation}
whose imaginary part reproduces Eq.~\eqref{eq:ImSigma} and defines the energy-dependent width $\Gamma(E) = -2\operatorname{Im}\Sigma_0(E)$. The physical resonance mass follows as $M_R(E) = M_0 + \operatorname{Re}\Sigma_0(E)$.

The model has three free parameters: $g_0$, $M_0$, and $\Lambda$. We determine them by a $\chi^2$ fit to the background-subtracted $\pi^{+}p$ elastic cross section of Ref.~\cite{Carter:1971tj} (main text Fig.~\ref{fig:FitCrosssection}), using the $P_{33}$ cross section formula~\cite{Haskins:1985xg,Giacosa:2021mbz}
\begin{equation}
\sigma(\Delta^{++}\rightarrow\pi^{+}p)=8\pi/k^2 \,(\operatorname{Im}T_{3/2}/|T_{3/2}|)^2.
\end{equation}
The best-fit parameters are
\begin{equation}
M_0 = 1327.69\;\text{MeV},\qquad
g_0 = 0.0261\;\text{MeV}^{-1/2},\qquad
\Lambda = 458.80\;\text{MeV}.
\end{equation}
With these parameters, the model reproduces the $\Delta(1232)$ pole $M_{\text{pole}}=1211.65 - i\,98.88/2\;\text{MeV}$ and the Breit-Wigner peak $M_R = 1231.19\;\text{MeV}$, consistent with PDG data~\cite{ParticleDataGroup:2024cfk} and confirming that the Friedrichs-Lee model faithfully describes the $\Delta$ resonance.

\subsection{Calculation in momentum space}
\label{subsec:FL-mom-space}

With the Friedrichs-Lee model parameters fixed, we incorporate a phenomenological Gaussian source to numerically evaluate the correlation function and verify the qualitative picture developed in Sec.~\ref{sec:LS-SP} in this supplement.
As in Sec.~\ref{sec:LS-SP}, we work in momentum space and compute each contribution in Eq.~\eqref{eq:C_decomposed}.

The coordinate-space Gaussian source,
$S(r) \propto \exp(-r^{2}/4R^{2})$, is Fourier transformed to momentum space:
\begin{equation}
\label{eq:S_mom}
S(\vct{k},\vct{k}') = \int d^{3}r\; e^{i(\vct{k}'-\vct{k})\cdot\vct{r}} S(r)
= \exp\!\bigl[-R^{2}(\vct{k}'-\vct{k})^{2}\bigr],
\end{equation}
where the overall normalization cancels in the correlation function and is omitted.
To isolate the $P$-wave component needed for the $\Delta(1232)$ resonance, we perform a Legendre expansion:
\begin{equation}
\label{eq:S_legendre}
S(\vct{k},\vct{k}') = \sum_{l=0}^{\infty} S_l(k,k')\,P_l(\hat{\vct{k}}\cdot\hat{\vct{k}}').
\end{equation}
The $P$-wave projection is the relevant channel and evaluates to
\begin{equation}
\label{eq:S1}
S_1(k,k') = \frac{3}{4k^{2}k'^{2}R^{4}}\,
\Bigl[ e^{-R^{2}(k+k')^{2}}\bigl(2kk'R^{2} + e^{4kk'R^{2}}(2kk'R^{2}-1) + 1\bigr) \Bigr].
\end{equation}

We account for the proton spin degrees of freedom via the source spin density matrix. For simplicity, in $pp$ collisions, we assume the source is unpolarized,
\begin{equation}
 \label{eq:source-spin}
 (\rho_S)_{s_z,s_z'} = \frac{1}{2}\,\delta_{s_z,s_z'}.
\end{equation}

Substituting the momentum-space Gaussian source [Eq.~\eqref{eq:S_mom}] and the Friedrichs-Lee $T$-matrix [Eq.~\eqref{eq:T_kk}] into Eqs.~\eqref{eq:C_int}--\eqref{eq:C_scat}, and performing the angular integrations together with the spin trace over unpolarized nucleons, we obtain the one-dimensional integral representations
\begin{align}
\label{eq:num-C_int}
C_{\text{int}}(p) &= \frac{2}{3}\times 2\operatorname{Re}
\int_{0}^{\infty} k'^{2}dk'\;
S_1(p,k')\, \frac{1}{E_p-E_{k'}+i\epsilon}\,
\frac{(g_0/m_\pi)^{2}\,k'p\,f(k')f(p)}{E_p-M_0-\Sigma_0(E_p)+i\epsilon},\\[4pt]
\label{eq:num-C_scat}
C_{\text{scat}}(p) &= \frac{2}{3}
\iint_{0}^{\infty} k^{2}dk\,k'^{2}dk'\;\\
&\qquad\frac{(g_0/m_\pi)^{2}\,kp\,f(k)f(p)}{E_p-M_0-\Sigma_0^{*}(E_p)-i\epsilon}\,
\frac{1}{E_p-E_k-i\epsilon}\,
S_1(k,k')\,
\frac{1}{E_p-E_{k'}+i\epsilon}\,
\frac{(g_0/m_\pi)^{2}\,k'p\,f(k')f(p)}{E_p-M_0-\Sigma_0(E_p)+i\epsilon}.\nonumber
\end{align}
The prefactor $2/3$ follows from the spin sum in the $P_{33}$ channel, $(2J+1)/[(2s+1)(2l+1)] = 4/(2\times 3) = 2/3$.

Applying the Sokhotski-Plemelj decomposition~\eqref{eq:SP} separates the on-shell and off-shell contributions. The on-off cross term $C_{\text{scat}}^{(\text{on-off})}$ of Eq.~\eqref{eq:C-sact-on-off} vanishes for the symmetric Gaussian source and is omitted.

\begin{align}
\label{eq:num-C-sact-on}
C_{\text{scat}}^{(\text{on-shell})}(p) &= \frac{2}{3}\,
\frac{\Gamma^{2}(E_p)/4}{(E_p-M_{\Delta}(E_p))^{2} + \Gamma^{2}(E_p)/4}\,
S_1(p,p),
\quad\text{(red line in the main text Fig.~\ref{fig:Corr-Apart})}\\[4pt]
\label{eq:num-C-int-on}
C_{\text{int}}^{(\text{on-shell})}(p) &= -\frac{2}{3}\,
\frac{\Gamma^{2}(E_p)/2}{(E_p-M_{\Delta}(E_p))^{2} + \Gamma^{2}(E_p)/4}\,
S_1(p,p),
\quad\text{(green line in the main text Fig.~\ref{fig:Corr-Apart})}\\[4pt]
\label{eq:num-C-scat-off}
C_{\text{scat}}^{(\text{off-shell})}(p) &= \frac{2}{3}\,
\frac{\Gamma(E_p)/2\pi\rho(E_p)}{(E_p-M_{\Delta}(E_p))^{2} + \Gamma^{2}(E_p)/4}
\iint_{0}^{\infty} k^{2}dk\,k'^{2}dk'\;\\
&\qquad\left(\frac{g_0}{m_\pi}\right)^{2} kk'f(k)f(k')
\left(\operatorname{P.V.}\frac{1}{E_p-E_k}\right)\,
S_1(k,k')\,
\left(\operatorname{P.V.}\frac{1}{E_p-E_{k'}}\right),\nonumber
\\[4pt]
&\text{(purple dashed line in the main text Fig.~\ref{fig:Corr-Apart})}\nonumber\\[4pt]
\label{eq:num-C-int-off}
C_{\text{int}}^{(\text{off-shell})}(p) &= \frac{2}{3}\,
\frac{E_p-M_{\Delta}(E_p)}{(E_p-M_{\Delta}(E_p))^{2} + \Gamma^{2}(E_p)/4}
\operatorname{P.V.}\int_{0}^{\infty} k'^{2}dk'\;
S_1(p,k')\,\frac{(g_0/m_\pi)^{2}\,pk'\,f(p)f(k')}{E_p-E_{k'}},
\\
&\text{(yellow dashed line in the main text Fig.~\ref{fig:Corr-Apart})}\nonumber
\end{align}

\subsection{Calculation in coordinate space}
\label{subsec:FL-coor-space}

To compare with experimental measurements, the Friedrichs-Lee model must be augmented with the Coulomb interaction $V_c(r)$ and the $\pi$-$p$ background strong-interaction potential $V_b(r)$ fitted by ALICE Collaboration~\cite{ALICE:2025aur}, which necessitates a coordinate space treatment. The Friedrichs-Lee parameters ($g_0$, $M_0$, $\Lambda$) fitted in momentum space are kept unchanged.

Including $V_c(r)$ and $V_b(r)$, the Schr\"odinger equation for the $P_{33}$ channel reads
\begin{equation}
 \label{eq:S-eq-full}
 (H_0+V_c(r)+V_b(r))R^{(P_{33})}_{J_z}(r)
 + \frac{1}{E_p-M_0}\int \tilde{g}_0(r')\,\tilde{g}_0^*(r)\,R^{(P_{33})}_{J_z}(r')\,r'^{2}dr'
 = E_p\,R^{(P_{33})}_{J_z}(r).
\end{equation}
The projection onto the $P_{33}$ channel follows from the angular-momentum structure of the vertex [Eq.~\eqref{eq:FL_vertex}]. Here $\tilde{g}_0(r)$ is the Fourier transform of the angular-momentum-independent part of the vertex $g_0\,(k/m_\pi)\,f(k)$,
\begin{equation}
\tilde{g}_0(r) = \frac{i\sqrt{\pi}}{\sqrt{8}}\,\frac{g_0}{m_\pi}\,\Lambda^{4}\,e^{-\Lambda r},
\end{equation}
and $R^{(P_{33})}_{J_z}(r)$ is the corresponding radial wave function. Since Eq.~\eqref{eq:S-eq-full} does not depend on $J_z$, we drop the subscript and write $R^{(P_{33})}(r)$.

In the absence of the resonance, the correlation function is given by the standard Koonin-Pratt formula
\begin{equation}
C^{(0)}(p) = \int d^{3}r\,S(r)\sum_{l}(2l+1)\,|R^{(0)}_l(r)|^{2},
\end{equation}
where $R^{(0)}_l(r)$ are the radial wave functions without the resonance, satisfying
\begin{equation}
 \label{eq:S-eq-other}
 (H_0+V_c(r)+V_b(r))\,R^{(0)}_l(r) = E_p\,R^{(0)}_l(r).
\end{equation}
Since the resonance affects only the $P_{33}$ partial wave, the full correlation function is obtained by replacing the $P_{33}$ component of $C^{(0)}$ with the resonant wave function $R^{(P_{33})}$:
\begin{equation}
 \label{eq:KP-full}
 \begin{aligned}
  C(p)&=\int d^{3}r\,S(r)\sum_{l}(2l+1)\,|R^{(0)}_l(r)|^{2}\\
      &\quad+\frac{2J+1}{(2s+1)(2l+1)}
        \int d^{3}r\,S(r)\,(2l+1)\,
        \Bigl(|R^{(P_{33})}(r)|^{2} - |R^{(0)}_l(r)|^{2}\Bigr)
        \Big|_{P_{33}},
 \end{aligned}
\end{equation}
where the subscript $P_{33}$ on the second term selects $l=1$, $J=3/2$. In short, the correlation function decomposes into a resonance-free background plus a $P_{33}$ resonance correction.

Using Eq.~\eqref{eq:KP-full} with the Gaussian source radii extracted by the ALICE Collaboration in different $m_T$ bins~\cite{ALICE:2025aur}, we obtain the theoretical curves shown in the main text Fig.~\ref{fig:baseline}.

\subsection{Comparison with experiment in \texorpdfstring{$E_{\text{c.m.}}$}{Ecm} space}
\label{subsec:FL-compExp}

\begin{figure}
 \includegraphics[width=1\linewidth]{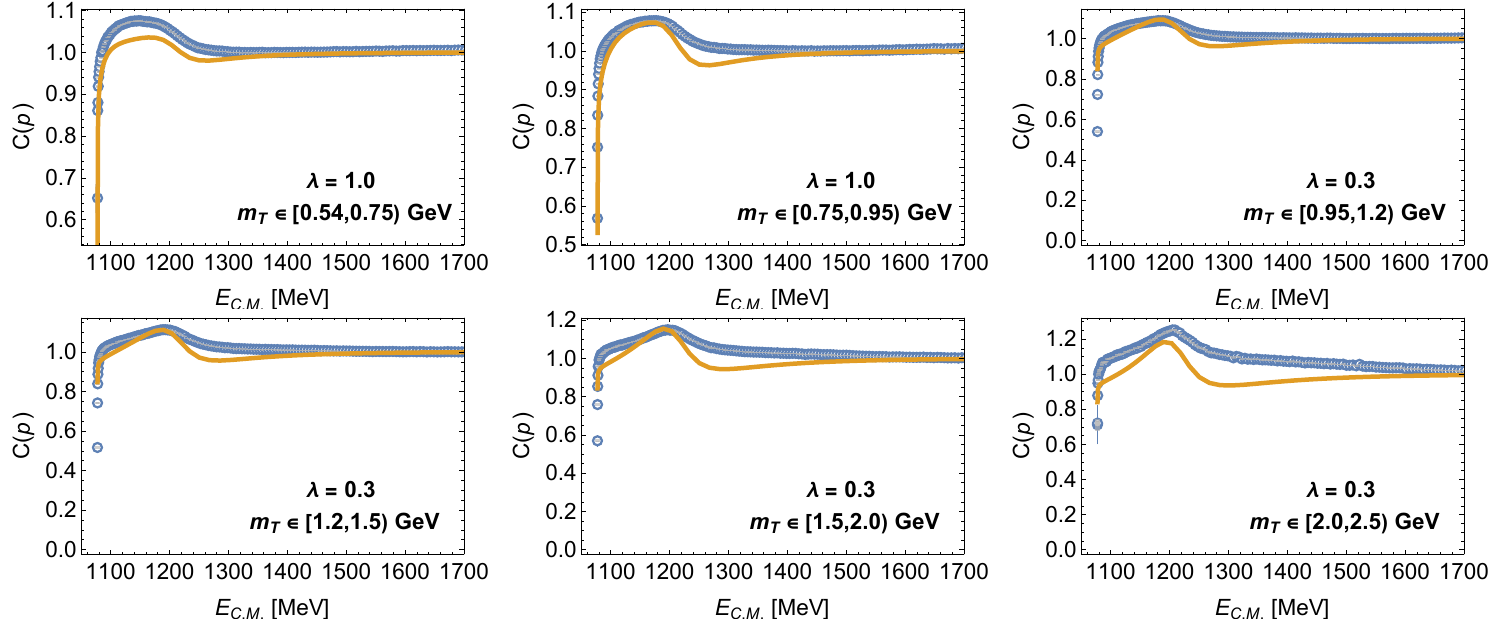}
 \caption{Same as the main text Fig.~\ref{fig:baseline}, but with the $x$-axis converted to $E_{\text{c.m.}}$. The experimental data are plotted using the relativistic energy-momentum relation, while the theoretical curves employ the non-relativistic approximation.}
 \label{fig:baseline-Ecm}
\end{figure}

The numerical calculations of Sec.~\ref{sec:FL-model} used non-relativistic kinematics $E_p = p^{2}/2\mu + m_p + m_\pi$, whereas experimentally the relative momentum and c.m.\ energy satisfy the relativistic relation $E_{\text{c.m.}} = \sqrt{m_p^{2}+p^{2}} + \sqrt{m_\pi^{2}+p^{2}}$.
This mismatch explains the different $\Delta$ mass reference lines in the main text Fig.~\ref{fig:Corr-Apart}: the solid (dashed) line marks the momentum corresponding to $M_{\Delta}$ under the relativistic (non-relativistic) relation.
Since the Friedrichs-Lee model parameters were fitted to cross-section data as a function of $E_{\text{c.m.}}$, it is natural to replot the main text Fig.~\ref{fig:baseline} with $E_{\text{c.m.}}$ as the $x$-axis.
The result is shown in Fig.~\ref{fig:baseline-Ecm}, where the theoretical curves agree more closely with the experimental data once the amplitude is tuned by the phenomenological parameter $\lambda$.

\subsection{Direct \texorpdfstring{$\Delta$}{Delta} contribution}
\label{subsec:FL-direct-Delta}

\begin{figure}
      \includegraphics[width=0.5\linewidth]{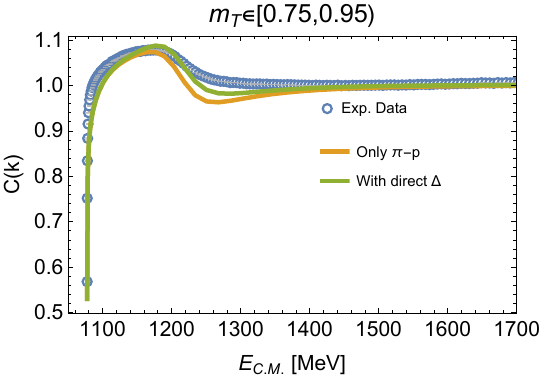}
      \caption{Contribution from directly produced $\Delta$ resonances surviving until freeze-out (green curve), which fills the high-momentum dip. The fitted $\Delta$ to $\pi$-$p$ yield ratio is $N_{\Delta}/N_{\pi p} \sim 5\times10^{-4}$ for $m_T\in[0.75,0.95)$~GeV.}
      \label{fig:Delta}
\end{figure}

The absence of the high-momentum dip in the experimental data can be attributed to directly produced $\Delta$ resonances that survive until kinetic freeze-out. As indicated by the second term in Eq.~\eqref{eq:P2_full}, this additional Breit-Wigner contribution compensates for the dip structure (Fig.~\ref{fig:Delta}). In the c.m.\ frame, the $S_{\mathrm{res}}$ term corresponds to the direct production of $\Delta$ resonances. After normalization via Eq.~\eqref{eq:corr-def}, this term reduces to the yield ratio of directly produced $\Delta$ to $\pi$-$p$ pairs, $N_{\Delta}/N_{\pi p}$, which is fitted to be $\sim 5\times10^{-4}$ for the $m_T\in[0.75,0.95)$~GeV bin shown in Fig.~\ref{fig:Delta}. As this ratio currently lacks independent support from existing measurements or model calculations, it is presented here as a supplementary observation and deferred from the main text.

\bibliography{04}